\documentclass{midl} % Include author names

\usepackage{mwe} % to get dummy images
\usepackage{float}

\jmlrproceedings{MIDL}{Medical Imaging with Deep Learning}
\jmlrpages{}
\jmlryear{2024}

\jmlrworkshop{Short Paper -- MIDL 2024 submission}
\jmlrvolume{-- Under Review}
\editors{Under Review for MIDL 2024}

\title[Contrastive learning unveils cortical folding pattern linked to prematurity]{Self-supervised contrastive learning unveils cortical folding pattern linked to prematurity}

 \midlauthor{\Name{Julien Laval} \Email{julien.laval@cea.fr}\\
  \Name{Denis Rivi\`ere} \Email{denis.riviere@cea.fr}\\
  \Name{Aymeric Gaudin} \Email{aymeric.gaudin@cea.fr}\\
  \Name{Vincent Frouin} \Email{vincent.frouin@cea.fr}\\
  \Name{Jessica Dubois}
  \Email{jessica.dubois@cea.fr}\\
  \Name{Andrea Gondova}
  \Email{andrea.gondova@etu.u-paris.fr
}\\
  \Name{Jean-François Mangin} \Email{jean-francois.mangin@cea.fr}\\
 \Name{Jo\"el Chavas}
 \Email{joel.chavas@cea.fr}\\
  \addr NeuroSpin, CEA Saclay, Université Paris-Saclay, France
  }

\begin{document}

\maketitle

\begin{abstract}
Brain folding patterns have been reported to carry clinically relevant information. The brain folds mainly during the last trimester of pregnancy, and the process might be durably disturbed by preterm birth. Yet little is known about preterm-specific patterns. In this work, we train a self-supervised model (SimCLR) on the UKBioBank cohort (21070 adults) to represent the right superior temporal sulcus (STS) region and apply it to sulci images of 374 babies from the dHCP database, containing preterms and full-terms, and acquired at 40 weeks post-menstrual age. We find a lower variability in the preterm embeddings, supported by the identification of a knob pattern, missing in the extremely preterm population. The code can be found on \href{https://github.com/neurospin-projects/2023_jlaval_STSbabies}{Github}.
\end{abstract}


\section{Introduction}

Brain folding can be described using sulci, which are the grooves separating surface humps, called gyri. Main sulci are common to all individuals, but they differ in shape. Recurrent shape patterns can be identified, and some have been reported to be correlated to pathologies~\cite{mellerio_power_2015,kabat_ogniskowa_2012}. Therefore, studying folding patterns can pave the way for better diagnosis and understanding of brain diseases. In particular, finding preterm-specific patterns could lead to a better understanding of their developmental trajectories and their susceptibility to neurodevelopmental diseases~\cite{hee_chung_neurodevelopmental_2020}. In this paper, we apply a self-supervised contrastive learning algorithm based on SimCLR~\cite{chen_simple_2020} to extract preterm-specific patterns.

\section{Methods} \label{methods}

\textbf{The dHCP database}: We process a subset of 374 subjects of the Developing Human Connectome Project (dHCP) database~\cite{edwards_developing_2022} containing subjects born between $23$ and $41$ post-menstrual weeks and classified according to their degree of prematurity (Table~\ref{table:age dHCP}). All images were acquired at $\sim40$ weeks post-menstrual using the same machine and protocol.
\begin{center}
\begin{table}[H]
\begin{center}
\begin{tabular}{ |c|c| } 
 \hline
 Degree of prematurity & number of subjects \\
 \hline\hline
 Extremely preterm ($<28$ weeks) & $19$ \\
 \hline
 Very preterm ($28 \leq$ weeks $< 32$) & $35$ \\
 \hline
 Moderate to late preterm ($32 \leq$ weeks $< 37$) & $43$ \\
 \hline
 full-term ($\geq37$ weeks) & $277$ \\
 \hline
\end{tabular}
\end{center}
\caption{Distribution of birth ages in dHCP, according to WHO classification}
\label{table:age dHCP}
\end{table}
\end{center}
\noindent\textbf{Inputs:} The starting points are structural MR images of the brain. They are processed through the BrainVisa Morphologist pipeline\footnote{\href{https://brainvisa.info}{https://brainvisa.info}} that skeletonizes a negative cast of the brain. It transforms the sulci into surfaces (3D objects of one-voxel width) following the middle of the folding. In practice, we use regional crops so that the algorithm can focus on specific sulci using the deep$\_$folding toolbox~\cite{chavas_unsupervised_2022}~\footnote{\href{https://github.com/neurospin/deep_folding}{github: neurospin/deep$\_$folding}}. Cortical skeleton images are affinely normalized in the Talairach space with a 2\~mm voxel size. This paper focuses on the right superior temporal sulcus (STS). It was chosen because its development spans the whole third trimester of pregnancy~\cite{dubois_mapping_2008}.

\noindent\textbf{Model parameters and evaluation:} \citet{gaudin_optimizing_2024} optimized a SimCLR pipeline applied to the cingular region of the brain, classifying the presence of the paracingular sulcus as a pretext task. They selected the backbone, the augmentations, and hyperparameters. The main augmentation is called "branch clipping". It removes branches of the skeleton until a given percentage of voxels is removed. No further optimization was performed in the present study because we want to evaluate the performance (thus the generalizability) of a model optimized on another brain region\footnote{The main parameters are the following:
backbone: ConvNet,
input size: (1, 22, 49, 46),
embedding size: 10,
number of trainable parameters: 452 172,
loss temperature: 0.1,
drop rate: 0.05,
branch clipping percentage: $40\%$,
batch size: 16,
lr: 0.0004,
weight decay: 5.0e-05,
max epochs: 250.}.
To assess the model performances, we apply a two-layer linear Support Vector Classifier to the model embeddings of dHCP for each prematurity group (the negative class being the full-terms) and measure the 3-fold cross-validation AUC (with class-based stratification).

\section{Experiments and results}

\begin{figure}[htb]
    \centering
    \includegraphics[width = 0.95\columnwidth]{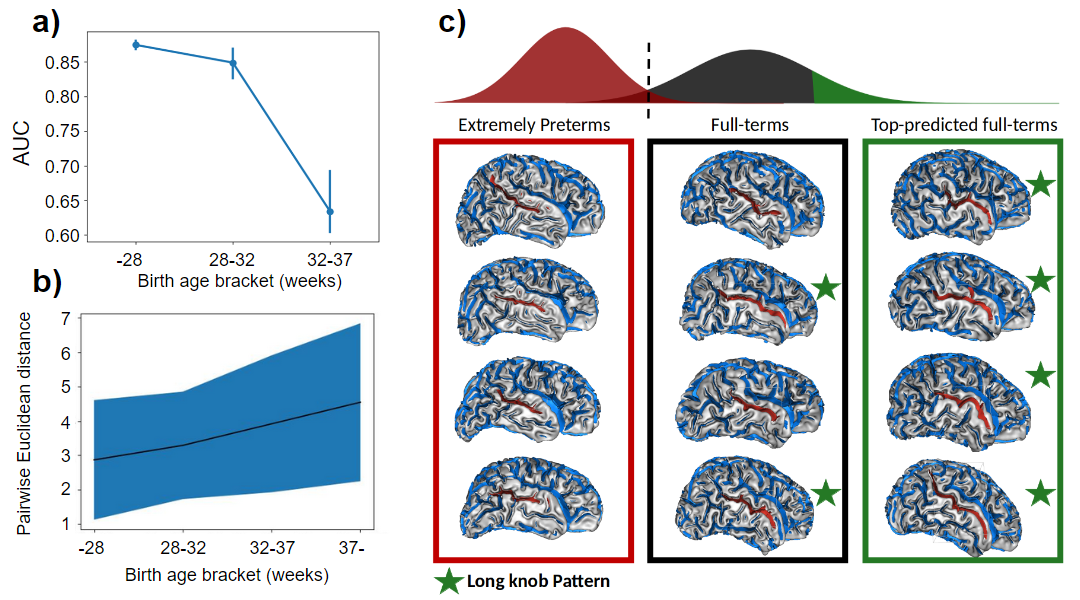}
    \caption{\textbf{a)} AUC between preterms and full-terms for each class of prematurity (averaged over $3$ trainings). \textbf{b)} Average pairwise Euclidean distance within each group in the latent space ($\pm2\sigma$ in blue). An F-test is performed to compute a p-value on the decrease of the variance~\textsuperscript{a} for each preterm category compared to full-term. $(,28)$: $\mathbf{p=0.01}$,\;$(28,32)$: $\mathbf{p=0.01}$,\;$(32,37)$: $p=0.1$. \textbf{c)} STS visualization (red) in whole brain context (sulci in blue) of dHCP subjects sampled based on the extremely preterm classification score. In each group, 19 subjects are sampled for the long knob pattern identification (marked in green). Only 4 are shown here, \citet{laval_self-supervised_2024} gives exhaustive visualization.}
    \label{fig:results}
    \small\textsuperscript{a} computed as the average squared Euclidean distance to the barycenter of the distribution.
\end{figure}

We train the model on the STS crops of 21070 adult brains from UkBioBank~\cite{sudlow_uk_2015}. Next, we compute the embeddings of the dHCP subjects and perform classification (\S\ref{methods}) to assess performances. We reach an AUC of $\mathbf{87.4(1.0)}$ for the extremely preterm class (Fig.~\ref{fig:results}a), confirming the ability of the model to capture information related to prematurity. Using UMAP~\cite{mcinnes_umap_2020}, we visualize the embedding space. The more premature the babies, the less variable they seem to be (see~\citet{laval_self-supervised_2024} for the UMAP). We quantify it in the 10-dimension latent space and find significant loss of variability for extremely and very preterms compared to full-terms (Fig.~\ref{fig:results}b). Then, we focus on the extremely preterms to characterize the variability drop visually. We select the top-predicted full-terms because they are more likely to contrast with preterms, and we identify an STS pattern shared by all top-predicted subjects, which is in stark contrast with typical extremely preterm STS: a long sulcus with a deep large knob at its anterior extremity. This pattern is then identified in at least $5/19$ randomly selected full-terms, which confirms quantitatively that it is missing in the extremely preterm population ($\mathbf{p=0.008}$ using Barnard's exact test, Fig.~\ref{fig:results}c).

\section{Conclusion}
In this work, using SimCLR, we identify an STS pattern absent from the extremely preterm data, suggesting a more unique developmental trajectory for preterms than full-terms. We showcase the transferability of the model optimized on adults in the cingular region to baby data in the STS region. Future work will be to repeat such analysis in all brain regions for the three classes of prematurity to uncover other regions altered by preterm birth.

% Acknowledgments---Will not appear in anonymized version
\midlacknowledgments{This work was funded by the ANR grants FOLDDICO and BABYTOUCH.}

\bibliography{biblio}

\end{document}